\documentclass[11pt]{article}
\usepackage{moriond,epsfig}
\usepackage{amssymb}

\def\be{\begin{equation}}
\def\ee{\end{equation}}
\def\bea{\begin{eqnarray}}
\def\eea{\end{eqnarray}}

\begin{document}
\vspace*{4cm}
\title{DIVERGENCES AND SYMMETRIES IN HIGGS-GAUGE UNIFICATION THEORIES}

\author{ C. BIGGIO }

\address{Institut de F\'\i sica d'Altes Energies (IFAE),
Universitat Aut\`onoma de Barcelona, \\
E-08193 Bellaterra (Barcelona), Spain}

\maketitle\abstracts{ In theories with extra dimensions the Standard
Model Higgs field can be identified with the internal components of
higher-dimensional gauge fields (Higgs-gauge unification). The
higher-dimensional gauge symmetry prevents the Higgs mass from
quadratic divergences, but at the fixed points of the orbifold this
symmetry is broken and divergences can arise if $U(1)$ subgroups are
conserved.  We show that another symmetry, remnant of the internal
rotation group after orbifold projection, can avoid the generation of
such divergences.}

\section{Introduction: Why Studying Higgs-Gauge Unification Theories?}
\label{sec:intro}

One of the possible motivations for studying Higgs-gauge unification
theories in extra
dimensions~\cite{Randjbar-Daemi:1982hi}$^\textrm{-}$\cite{Scrucca:2003ut}
is the so-called little hierarchy problem.\cite{Giudice:2003nc} If one
considers the Standard Model (SM) as an effective theory valid up to a
certain scale $\Lambda_{SM}$ and calculates the radiative corrections
to the Higgs mass, he finds that these diverge quadratically and that,
in order to avoid a fine-tuning of the parameters, $\Lambda_{SM}$ must
be smaller than 1 TeV. This means that new physics must enter into the
game at the scale of 1 TeV to regularize the ultraviolet behaviour. On
the other hand this new physics can be parametrized by adding to the
SM lagrangian non-renormalizable operators suppressed by powers of
$\Lambda_{LH}$. From the non-observation of dimension-six four-fermion
operators at LEP a lower limit on $\Lambda_{LH}$ of 5-10 TeV has been
derived.\cite{PDG} This one order of magnitude discrepancy between the
theoretically required upper limit and the experimental lower limit is
called little hierarchy problem.

Up to now the best solution to the little (and grand) hierarchy
problem is supersymmetry (SUSY). In the supersymmetric extensions of
the SM quadratic divergences are absent so that the supersymmetric
model can be extended up to $M_{Pl}$ without the need of other new
physics: this solves the hierarchy problem. The $\Lambda_{SM}$ is now
identified with the mass of the supersymmetric particles, which can be
$O(1\textrm{ TeV})$. Moreover, if R-parity is conserved, it induces a
suppression in the loop corrections to four-fermions operators which
results in the relation $\Lambda_{LH}\sim 4\pi \Lambda_{SM}$, that
precisely solves the little hierarchy problem. However SUSY has not
yet been discovered and this reintroduces a small amount of
fine-tuning in the theory. For this and other reasons we think it can
be worthwhile looking for alternative solutions to the little
hierarchy problem.

One alternative solution is given by the so-called Higgs-gauge
unifications theories, in the context of theories in extra dimensions
compactified on orbifolds. If we consider a gauge field in $D$
dimensions, its components can be split into two parts, according to
the transformation properties under the four-dimensional ($4D$)
Lorentz group: $A^A_M = (A^A_\mu , A^A_i)$, where $A^A_\mu$ is a $4D$
Lorentz vector while $A^A_i$ are $4D$ Lorentz scalars. The latter can
be identified with the Higgs fields and they can acquire a
non-vanishing vacuum expectation value through the Hosotani
mechanism.\cite{Hosotani:1983xw} The good feature of this kind of
constructions is that the Higgs mass in the bulk is protected from
quadratic divergences by the higher-dimensional gauge invariance and
only finite corrections $\propto (1/R)^2$, where $R$ is the
compactification radius, can appear. So the picture is the following:
we have the $4D$ SM valid up to the scale $\Lambda_{SM}\sim 1/R$ that
can be $O(1\textrm{ TeV})$, then we have a non-renormalizable
$D$-dimensional theory valid up to a certain scale $\Lambda_D$ which
can be greater or equal to $10\textrm{ TeV}$ and then we have the
ultraviolet completion of the theory. As we see also in this case the
little hierarchy problem is solved.

In this talk we will deal with some specific features of these
theories and, in particular, we will discuss how the mass protection
given by the higher-dimensional gauge invariance can be spoiled at the
fixed points of the orbifold and under which conditions it can be
restored.

\section{Symmetries at the Fixed Points of an Orbifold and Allowed Localized Terms}
\label{sec:main}

\subsection{Gauge Theories on Orbifolds}
\label{sub:gaugeteo}

We begin by considering a gauge theory coupled to fermions in a
$D$-dimensional ($D=d+4>4$) space-time parametrized by coordinates
$x^M=(x^\mu, y^i)$ where $\mu=0,1,2,3$ and $i=1,\dots ,d$. The
lagrangian is
\begin{equation} 
{\mathcal L}_{D}=-\frac{1}{4}
{F}_{MN}^A{F}^{AMN}
+i{\overline \Psi}\Gamma_D^M D_M{\Psi},
\label{bulk-lagr}
\end{equation}
with $F^A_{MN}=\partial_M A_N^A - \partial_N A_M^A - g f^{ABC} A_M^B
A_N^C$, $D_M=\partial_M-igA_M^AT^A$ and where $\Gamma_D^M$ are the
$\Gamma$-matrices corresponding to a $D$-dimensional space-time. This
lagrangian is invariant under the gauge group $\mathcal{G}$ and of
course under the $D$-dimensional Lorentz group $SO(1,D-1)$. 

\begin{figure}
\begin{center}
\psfig{figure=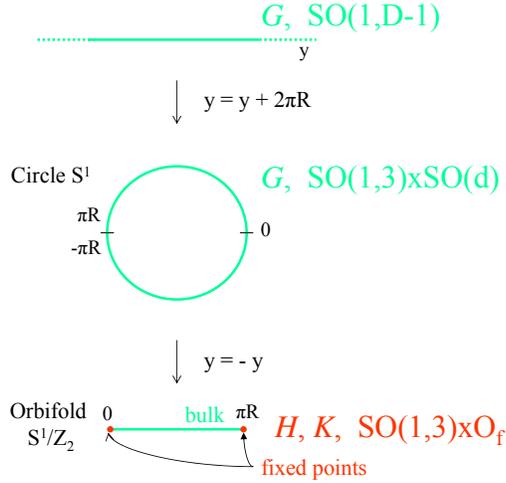,height=8cm}
\caption{The orbifold $S^1/Z_2$: construction and symmetry breaking pattern.
\label{disegno}}
\end{center}
\end{figure}

Now we compactify the extra dimensions on an orbifold.\cite{orbifold}
Firstly we build up a $d$-dimensional torus $T^d$ by identifying
$(x^\mu, y^i)$ with $(x^\mu, y^i + u^i)$, with $\vec{u}$ belonging to
a $d$-dimensional lattice $\Lambda^d$. Then we act on the torus with
the element $k$ of the group $\mathbb G$ generated by a discrete
subgroup of $SO(d)$ that acts crystallographically on the torus
lattice and by discrete shifts that belong to it. The orbifold is
finally defined by the identification $(x^\mu, y^i)=(x^\mu, (P_k\
\vec{y})^i + u^i)$, where $P_k$ is the rotation associated to the
element $k\in \mathbb G$ . This group acts non-freely on the torus,
i.e.~it leaves some points invariant: these are called fixed
points. The construction of the orbifold $S^1/Z_2$ in five dimensions
is depicted in Fig.~\ref{disegno}: by identifying $y$ with $y+2 n \pi
R$ (with $n$ integer) we are left with a segment of length $2\pi R$
and the extrema identified which can be represented by a circle; then
by identifying $y$ with $-y$ we end up with a segment of length $\pi
R$, with two fixed points in $y_f=0,\pi R$, which is the orbifold.

Now that we have defined the orbifold we have to specify how it acts
on the fields. If $\phi_{\mathcal R}$ is a generic field transforming
as an irreducible representation ${\mathcal R}$ of the gauge group
$\mathcal G$, then the orbifold action is defined by
\begin{equation}
k\cdot\phi_{\mathcal R}(y)
=\lambda^k_{\mathcal R}\otimes\mathcal P^k_\sigma\,
\phi_{\mathcal R}(k^{-1}\cdot y)
\label{orbfield}
\end{equation}
where $\lambda^k_{\mathcal R}$ is acting on gauge and flavor indices
and $\mathcal P^k_\sigma$, where $\sigma$ refers to the field spin, on
Lorentz indices. Splitting the action of the orbifold in this way is
particularly useful since while $\mathcal P^k_\sigma$ is fixed by
requiring the invariance of the lagrangian under this transformation
(in particular we obtain $\mathcal P_0^k=1$ for scalar fields and
$\mathcal P_1^k=P_k$ for gauge fields), $\lambda^k_{\mathcal R}$ is
unconstrained and it can be used to break symmetries.\footnote{For a
review on symmetry breaking on orbifolds see for instance Ref.~[9] and
references therein.}

Now we consider the gauge symmetry breaking realized by the orbifold
at the generic fixed point $y_f$. First of all we have to understand
why we are looking precisely there. The reason is that new lagrangian
terms localized at the fixed points can be generated by bulk radiative
corrections~\cite{ggh} if they are compatible with the existing
symmetries. Since we are interested in the stability of the Higgs mass
under radiative corrections, the knowledge of these symmetries can
tell us if our theory is stable or not.

In general the orbifold action breaks the gauge group in the bulk
$\mathcal G=\{T^A\}$ to a subgroup $\mathcal H_f=\{T^{a_f}\}$, at the
fixed point $y_f$, defined by the generators of $\mathcal G$ which
commute with $\lambda^k_{\mathcal R}$, i.e.~$[\lambda^k_{\mathcal
R},T^{a_f}_{\mathcal R}]=0$. This condition must be satisfied by any
irreducible representation $\mathcal R$ of $\mathcal G$.  The symmetry
breaking pattern also defines which fields are non-zero at $y_f$. In
this case they will be $A^a_\mu$, which are the gauge bosons of the
unbroken gauge group $\mathcal H_f$, and $A^{\hat{a}}_i$, for some $i$
and $\hat{a}$, with zero modes, plus some derivatives of non-invariant
fields without zero modes. Since also the parameters which define the
gauge transformation transform under the orbifold action, the
derivatives of some of them which are invariant define a set of local
transformations that are called $\mathcal
K$-transformations.\cite{vonGersdorff:2002us} Eventually at the fixed
point $y_f$ there are two symmetries remnant of the original gauge
symmetry $\mathcal G$: $\mathcal H_f$ and $\mathcal K$.  All this is
summarized in Fig.~\ref{disegno}.

\subsection{Allowed Localized Lagrangian Terms}
\label{sub:symmfix}

Now that we have discussed the main features of gauge symmetry
breaking on orbifold, we are ready to write down the most general $4D$
effective lagrangian. This is given by the integral over the extra
coordinates of the $D$-dimensional lagrangian plus the terms localized
at the fixed points which are compatibles with the symmetries:
\begin{equation}
\mathcal L_4^{eff}=\int d^d y \bigl[
\mathcal L_D+\sum_{f}\delta^{(d)}(y-y_f)\,\mathcal L_f \bigr].
\label{eff-lag}
\end{equation}
How are exactly these $\mathcal L_f$? Before answering to this
question we have to list the symmetries holding at the fixed
points. These are the orbifold group [$\mathbb G_f$], the $4D$ Lorentz
group [$SO(1,3)$], the residual gauge group [$\mathcal H_f$] and the
residual local symmetry [$\mathcal K$].

The $\mathcal K$-symmetry is very important since it forbids the
appearance of direct mass terms like $\Lambda^2 A^{\hat{a}}_i
A^{\hat{b}}_j$ in the case in which $A^{\hat{a}}_i$ is $\mathbb
G_f$-invariant.  Anyway the previously listed symmetries allow
localized terms as $(F^a_{\mu\nu})^2$, which corresponds to a
localized kinetic term for $A_\mu^a$, and
$F^a_{\mu\nu}\tilde{F}^{a\mu\nu}$ which is a localized
anomaly. Moreover if for some $(i,j)$ $F^a_{ij}$ and $A_i^{\hat a}$
are orbifold invariant (this is model-dependent), $(F^a_{ij})^2$ and
$(F^{\hat a}_{i\mu})^2$ are also allowed, giving rise respectively to
localized quartic couplings and kinetic terms for $A_i^{\hat a}$.  All
these operators are dimension-four, that is they renormalize
logarithmically. However if $\mathcal H_f$ contains a $U(1)$ factor
\begin{equation}
F_{ij}^\alpha=\partial_i A_j^\alpha-\partial_jA_i^\alpha-gf^{\alpha \hat
b \hat c}A_i^{\hat b}A_j^{\hat c},
\label{fij}
\end{equation}
where $\alpha$ is the $U(1)$ quantum number, is invariant under all
the above discussed symmetries and can be generated by bulk radiative
corrections at the fixed points. This means that we expect both a
tadpole for the derivatives of odd fields and a mass term for the even
fields. Since these operators are dimension-two, their respective
renormalizations will lead to quadratic divergences, making the theory
ultraviolet-sensitive.

Apart from the $5D$ case where the term $F_{ij}$ does not exist, for
$D\ge 6$ it does and its generation has been confirmed by direct
computation in 6D orbifold
field~\cite{vonGersdorff:2002us}$^\textrm{-}$\cite{Scrucca:2003ut} and
10D string~\cite{GrootNibbelink:2003gb} theories.  Of course if these
divergent localized mass terms were always present, Higgs-gauge
unification theories would not be useful in order to solve the little
hierarchy problem.  One way out can be that local tadpoles vanish
globally, but this requires a strong restriction on the bulk fermion
content.\cite{Scrucca:2003ut} A more elegant and efficient solution,
based on symmetry arguments, has been presented in Ref.~[12] and will
be discussed in the following.

\subsection{The Residual $O_f$ Symmetry}
\label{sub:of}

When we discussed the symmetry breaking induced by the orbifold, we
did not consider the $D$-dimensional Lorentz group.  When
compactifying a $d$-dimensional space to a smooth Riemannian manifold
(with positive signature), at each point a tangent space can be
defined and the orthogonal transformations acting on it form the group
$SO(d)$.\cite{Witten} When the orbifold group acts on the manifold, in
the same way as the gauge group $\mathcal G$ is broken down to
$\mathcal H_f$, the internal rotation group $SO(d)$ is broken down to
a subgroup $\mathcal O_f$ defined by the generators of $SO(d)$ which
commutes with $\mathcal P^k_\sigma$, i.e.~$[\mathcal
P^k_\sigma,\mathcal O_f]=0$. This means that the original Lorentz
group $SO(1,D-1)$ is firstly broken down to $SO(1,3)\otimes SO(d)$
(where $SO(d)$ must be understood as acting on the tangent space) by
the smooth compactification and then it is definitively broken down to
$SO(1,3)\otimes \mathcal O_f$ by the orbifold action. All this is
outlined in Fig.~\ref{disegno}.

We have then identified an additional symmetry that the lagrangian
$\mathcal L_f$ at the fixed point $y_f$ must conserve. Summarizing,
the invariances that we have to take into account are the following:
4D Lorentz invariance [$SO(1,3)$], invariance under the action of the
orbifold group [$\mathbb G_f$], usual 4D gauge invariance [$\mathcal
H_f$], remnant of the bulk gauge invariance [$\mathcal K$] and remnant
of the invariance under rotations of the tangent space [$\mathcal
O_f$]. Now the question is: can this $\mathcal O_f$ forbid the
appearance of the tadpole (or, equivalently, of the divergent mass
term for the Higgs)?

If $\mathcal O_f$ contains among its factors at least one $SO(2)$,
then a corresponding Levi-Civita tensor $\epsilon^{ij}$ exists, such
that the lagrangian term $\epsilon^{ij} F_{ij}^{(\alpha)}$ is also
$\mathcal O_f$-invariant. In this case tadpoles are allowed. On the
other side, if $\mathcal O_f$ is given by a product of $SO(p_i)$ with
$p_i > 2\ \forall i$, then the Levi-Civita tensor has $p_i$ indices
and only invariants constructed using $p_i$-forms are allowed. Since
$F_{ij}^\alpha$ has two indices this means that in this case tadpoles
are not allowed. We have then found a sufficient condition for the
absence of localized tadpoles which precisely is that the smallest
internal subgroup factor be $SO(p)$ with $p>2$.

Evidently $\mathcal O_f$ is orbifold-dependent; in Ref.~[12] we
analyzed the case of the orbifold $T^d/Z_N$ for $d$ even. In this case
the generator of the orbifold group is given by $P_N={\rm
diag}(R_1,\dots,R_{d/2})$, where $R_i$ is the discrete rotation in the
$(y_{2i-1},y_{2i})$-plane.  If $\mathbb Z_{N_f}$ is the orbifold
subgroup which leaves invariant the point $y_f$, it can be shown that
if $N_f>2$ then $\mathcal O_f=\bigotimes_{i=1}^{d/2}SO(2)_i$, where
$SO(2)_i$ is the $SO(2)\subseteq SO(d)$ that acts on the
$(y_{2i-1},y_{2i})$-subspace. Then in every subspace $\epsilon^{IJ}$
exists and we expect a tadpoles appearance at the fixed points $y_f$
of the form
\begin{equation}
\sum_{i=1}^{d/2}\mathcal
C_i\sum_{I,J=2i-1}^{2i}\epsilon^{IJ}F^\alpha_{IJ}\;\delta^{(d/2)}(y-y_f).
\label{tadpolos}
\end{equation}
On the contrary if $N_f=2$ then the generator of the orbifold subgroup
is the inversion $P_2=-\mathbf{1}$ that obviously commutes with all
the generators of $SO(d)$ so that we have $\mathcal O_f=SO(d)$. In
this case the Levi-Civita tensor is $\epsilon^{i_1\dots i_d}$ and only
a $d$-form can be generated linearly in the localized
lagrangian. Therefore tadpoles are only expected in the case of $d=2$
$(D=6)$. This last comment also apply to the case of $\mathbb Z_2$
orbifolds in arbitrary dimensions (even or odd), since the orbifold
generator is always $P=-\mathbf{1}$ and then the internal rotation
group is always $\mathcal O_f=SO(d)$.  In Ref.~[12] we explicitly
checked this result at one- and two-loops for the orbifold $T^d/Z_2$
for any $D$.

\section{Conclusions}
\label{sec:conclu}

In orbifold field theories the SM Higgs field can be identified with
the internal components of gauge fields. Then the higher-dimensional
gauge invariance prevents the Higgs from acquiring a quadratically
divergent mass term in the bulk, while at the fixed points a remnant
of bulk gauge symmetry after symmetry breaking forbids the appearance
of direct mass terms. Still, if the residual gauge symmetry contains a
$U(1)$ factor, the corresponding field strength for the $4D$ scalar
fields is invariant under the orbifold action, the $4D$ Lorentz
symmetry, the residual gauge invariance and the residual local
symmetry, so that it can be radiatively generated at the fixed
points. This is a dimension-two operator and gives rise to a
quadratically divergent mass for the Higgs. However we showed that
another symmetry must be considered. Indeed, when compactifying on an
orbifold, the internal rotation group acting on the tangent space that
can be defined at each point of a smooth manifold is broken down at
the fixed points, since there a tangent space cannot be defined. How
the breaking is realized depends on the particular orbifold but in
general a group $\mathcal O_f$, subgroup of the internal rotation
group, will survive and then it shall be taken into account when
looking for lagrangian terms that can be radiatively
generated. Actually this residual symmetry can forbid the appearance
of dangerous divergent terms. Indeed if $\mathcal O_f$ contains among
its factors at least one $SO(2)$, then a Levi-Civita tensor
$\epsilon^{ij}$ exists and the previously mentioned invariant field
strength will be generated at the considered fixed point in the form
$\epsilon^{ij}F^{(\alpha)}_{ij}$. On the contrary if $\mathcal O_f$ is
given by a product of $SO(p_i)$ with each $p_i>2$, then only
invariants constructed with $p_i$-forms can be generated and our
dangerous term will not appear. What we have found is then another
sufficient condition for the absence of localized tadpoles. Also we
have shown that in the case of the orbifolds $T^d/Z_N$ ($N>2$, $d$
even) $\mathcal O_f$ is a product of $SO(2)$ groups (at least at the
$Z_N$-fixed points) and then divergent terms will always be
allowed. On the other side for the orbifolds $T^d/Z_2$ (any $d$)
$\mathcal O_f$ always coincides with the whole $SO(d)$ and then
tadpoles will never appear for $d>2$ ($D>6$).

\section*{Acknowledgments}

I would like to thank Mariano Quir\'os for the pleasant collaboration
on which this talk is based. I also would like to thank the organizers
of the ``XL Rencontres de Moriond'' for the enjoyable atmosphere of
this interesting conference.

\end{document}